\documentclass{elsart}
\usepackage{graphics}
\usepackage{graphicx}
\usepackage{epstopdf}
\usepackage{epsfig}
\usepackage{amssymb}
\usepackage{makeidx}
\usepackage{amsfonts}
\usepackage{amstext}
\usepackage{amsmath}
\usepackage{amsbsy}
\usepackage{rotating}

\newcommand{\po}{\mbox{$\pi^0$}}

\renewcommand{\deg}{\mbox{$^\circ$}}
\def\A{\kern+.6ex\lower.42ex\hbox{$\scriptstyle \iota$}\kern-1.20ex a}
\def\E{\kern+.5ex\lower.42ex\hbox{$\scriptstyle \iota$}\kern-1.10ex e}

\begin{document}
\begin{frontmatter}

\title{Energy reconstruction of electromagnetic
showers from $\pi^0$ decays with the ICARUS
T600 Liquid Argon TPC}
%
{\large\sf ICARUS Collaboration}\\
\author[ITP-Wroclaw]{\small A. Ankowski}{,}
\author[LNGS]{\small M. Antonello}{,}
\author[LNGS]{\small P.G. Aprili}{,}
\author[LNGS]{\small F. Arneodo}{,}
\author[ETHZ]{\small A. Badertscher}{,}
\author[Padova]{\small B. Baibussinov}{,}
\author[UPadova,Padova]{\small M.Baldo-Ceolin}{,}
\author[Milano]{\small G. Battistoni }{,}
\author[UPavia,Pavia]{\small P. Benetti}{,}
\author[Pavia]{\small R. Brunetti}\footnote{Now at INFN - Sezione di Torino,
Torino, Italy}{,}
\author[Granada]{\small A. Bueno}{,}
\author[Pavia]{\small E.Calligarich}{,}
\author[UPavia,Pavia]{\small M. Cambiaghi}{,}
\author[UAquila,LNGS]{\small N. Canci}{,}
\author[UNapoli,Napoli]{\small F. Carbonara}{,}
\author[Granada]{\small M.C. Carmona}{,}
\author[UAquila,LNGS]{\small F. Cavanna}{,}
\author[CERN]{\small P. Cennini}{,}
\author[UPadova,Padova]{\small S. Centro}{,}
\author[PoliMI,Milano]{\small A. Cesana}{,}
\author[Krakow,Padova]{\small K. Cie\'slik}{,}
\author[UCLA]{\small D. Cline}{,}
\author[Napoli]{\small A.G. Cocco}{,}
\author[Krakow]{\small A. D\A browska}{,}
\author[UPavia,Pavia]{\small R. Dolfini}{,}
\author[UPadova,Padova]{\small C. Farnese}{,}
\author[UPadova,Padova]{\small A. Fava}{,}
\author[CERN]{\small A. Ferrari}{,}
\author[UNapoli,Napoli]{\small G. Fiorillo}{,}
\author[UAquila,LNGS]{\small S. Galli}{,}
\author[Napoli]{\small V. Gallo}{,}
\author[Granada]{\small D. Garcia-Gamez}{,}
\author[UPadova,Padova]{\small D. Gibin}{,}
\author[UPavia,Pavia]{\small A. Gigli Berzolari}{,}
\author[ITP-Wroclaw]{\small K. Graczyk}{,}
\author[Padova]{\small A. Guglielmi}{,}
\author[IP-Katowice]{\small J. Holeczek}{,}
\author[IEP-UWarsaw]{\small D. Kie\l{}czewska}{,}
\author[IP-Katowice]{\small J. Kisiel}{,}
\author[Soltan-Warsaw]{\small T. Koz\l{}owski}{,}
\author[Soltan-Warsaw]{\small J. \L{}agoda}{,}
\author[Milano]{\small M. Lantz}\footnote{Now at Chalmers University, Sweden}{,}
\author[Granada]{\small J. Lozano}{,}
\author[LNF]{\small G. Mannocchi}{,}
\author[Krakow]{\small M. Markiewicz}{,}
\author[Granada]{\small A. Martinez de la Ossa}{,}
\author[Pavia]{\small F. Mauri}$^\dagger${,}
\author[Granada]{\small A. J. Melgarejo}{,}
\author[Pavia]{\small A. Menegolli\corauthref{cor}}{,}
\corauth[cor]{Corresponding author}
\author[Padova]{\small G. Meng}{,}
\author[Soltan-Warsaw]{\small P. Mijakowski}{,}
\author[Pavia]{\small C. Montanari}{,}
\author[UAquila,LNGS]{\small G. Piano Mortari}{,}
\author[Milano]{\small S. Muraro}{,}
\author[Granada]{\small S. Navas}{,}
\author[UCLA]{\small S. Otwinowski}{,}
\author[LNGS]{\small O. Palamara}{,}
\author[Soltan-Warsaw]{\small T. J. Palczewski}{,}
\author[LNF]{\small L. Periale}{,}
\author[UPavia,Pavia]{\small A. Piazzoli}{,}
\author[LNF]{\small P. Picchi}{,}
\author[Padova]{\small F. Pietropaolo}{,}
\author[AGH-Krakow]{\small W. P\'o\l{}ch\l{}opek}{,}
\author[IEP-UWarsaw]{\small M. Posiada\l{}a}{,}
\author[Pavia]{\small M. Prata}\footnote{Now at LENA, Universit\'a di Pavia, Italy}{,}
\author[Soltan-Warsaw]{\small P. Przew\l{}ocki}{,}
\author[Pavia]{\small A. Rappoldi}{,}
\author[Pavia]{\small G.L. Raselli}{,}
\author[Soltan-Warsaw]{\small E. Rondio}{,}
\author[Pavia]{\small M. Rossella}{,}
\author[ETHZ]{\small A. Rubbia}{,}
\author[LNGS]{\small C. Rubbia}{,}
\author[Milano]{\small P. Sala}{,}
\author[UPavia,Pavia]{\small N. Scannicchio}{,}
\author[Milano]{\small A. Scaramelli}{,}
\author[LNGS]{\small E. Segreto}{,}
\author[Pisa]{\small F. Sergiampietri}{,}
\author[ITP-Wroclaw]{\small J. Sobczyk}{,}
\author[Krakow]{\small D. Stefan}{,}
\author[Soltan-Warsaw]{\small J. Stepaniak}{,}
\author[IRE-UWarsaw]{\small R. Sulej}{,}
\author[Krakow]{\small M. Szarska}{,}
\author[IP-Katowice]{\small T. Szeglowski}{,}
\author[Soltan-Warsaw]{\small M. Szeptycka}{,}
\author[PoliMI,Milano]{\small M. Terrani}{,}
\author[UPadova]{\small F. Varanini}{,}
\author[Padova]{\small S. Ventura}{,}
\author[Pavia]{\small C. Vignoli}{,}
\author[Krakow]{\small T. W\A cha\l{}a}{,}
\author[UCLA]{\small H. Wang}{,}
\author[Krakow]{\small A. Zalewska}
\clearpage
\address[UAquila]{Universit\`a dell'Aquila, L'Aquila, Italy}
\address[LNGS]{INFN - Laboratori Nazionali del Gran Sasso, Assergi (AQ), Italy}
\address[CERN]{CERN, Geneva, Switzerland}
\address[Milano]{INFN - Sezione di Milano, Milano, Italy}
\address[PoliMI]{CESNEF - Politecnico di Milano, Milano, Italy}
\address[UNapoli]{INFN - Sezione di Napoli, Napoli, Italy}
\address[Napoli]{Universit\'a Federico II  di Napoli, Napoli, Italy}
\address[Padova]{INFN - Sezione di Padova, Padova, Italy}
\address[UPadova]{Universit\'a di Padova, Padova, Italy}
\address[Pisa]{INFN - Sezione di Pisa, Pisa, Italy}
\address[UPavia]{Universit\'a di Pavia, Pavia, Italy}
\address[Pavia]{INFN - Sezione di Pavia, Pavia, Italy}
\address[LNF]{INFN - Laboratori Nazionali di Frascati, Frascati (Roma), Italy}
\address[UCLA]{UCLA - University of California, Los Angeles, California, USA}
\address[IP-Katowice]{Institute of Physics - University of Silesia, Katowice, Poland}
\address[Krakow]{H. Niewodnicza\'nski Institute of Nuclear Physics PAN, Krak\'ow, Poland}
\address[AGH-Krakow]{AGH University of Science and Technology, Krak\'ow, Poland}
\address[IEP-UWarsaw]{Institute of Experimental Physics - University of Warsaw, Warsaw, Poland}
\address[IRE-UWarsaw]{Institute of Radioelectronics, Warsaw University of Technology, Warsaw, Poland}
\address[Soltan-Warsaw]{A. So\l{}tan Institute for Nuclear Studies, Warsaw, Poland}
\address[ITP-Wroclaw]{Institute of Theoretical Physics, University of Wroc\l{}aw, Wroc\l{}aw, Poland}
\address[ETHZ]{Institute for Particle Physics, ETH H\"onggerberg, Z\"urich, Switzerland}
\address[Granada]{Universidad de Granada and CAFPE, Granada, Spain}
\begin{abstract}
We discuss the ICARUS T600 detector capabilities in electromagnetic
shower reconstruction through the analysis of a sample of 212
events, coming from the 2001 Pavia surface test run, of hadronic
interactions leading to the production of $\pi^{0}$ mesons. Methods
of shower energy and shower direction measurements were developed
and the invariant mass of the photon pairs was reconstructed. The
($\gamma$,$\gamma$) invariant mass was found to be consistent with
the value of the $\pi^0$ mass. The resolution of the reconstructed
$\pi^0$ mass was found to be equal to 27.3\%.
An improved analysis, carried out in order to clean the full event sample
from the events measured in the crowded environment, mostly due to the trigger
conditions, gave a $\pi^0$ mass resolution of 16.1\%, significantly better than the
one evaluated for the full event sample. The trigger requirement of the
coincidence of at least four photomultiplier signals favored the selection of
events with a strong pile up of cosmic ray tracks and interactions. Hence a
number of candidate $\pi^0$ events were heavily contaminated by other tracks and
had to be rejected. Monte Carlo simulations of
events with $\pi^0$ production in hadronic and neutrino interactions
confirmed the validity of the shower energy and shower direction
reconstruction methods applied to the real data.
\end{abstract}

\end{frontmatter}

\section{Introduction}

The Liquid Argon (LAr) Time Projection Chamber (TPC) is an option
for the next generation of very large mass neutrino detectors. The
LAr TPC was proposed by C.Rubbia~\cite{Rubbia_LAr} in 1977 and,
after many years of extensive R\&D studies the
T600 ICARUS detector~\cite{t600paper} has been constructed. The
read-out of ionization electrons traveling in a uniform electric
field provides the spacial reconstruction of an event and allows
for calorimetric measurement of the energy release. In year 2001,
the T600 ICARUS detector was extensively tested in Pavia (Italy),
with an exposure to cosmic rays at the Earth surface. During over
100 days of continuous data taking, about 30000 cosmic ray events
have been collected. In a series of papers the ICARUS
Collaboration presented the following results from the T600
detector test run: observation of long ionizing
tracks~\cite{long_track}, LAr purity~\cite{purity}, electron
recombination in LAr~\cite{e_recombination} and measurement of through-going
particle momentum by means of multiple scattering~\cite{momentum}.
The analysis of the sample of stopping $\mu$ events led to the measurement
of the $\mu$ decay spectrum, and allowed to determine the energy resolution of the
T600 ICARUS detector for the measurement of electrons with energy
lower than about 50 MeV~\cite{muon_decay}.\\
The ICARUS T600 detector is at present in the final phases of
re-assembling in the underground Hall B at INFN Laboratori Nazionali del Gran Sasso
(LNGS), Italy, and its commissioning is foreseen to start
within the year 2008.\\
In this paper we discuss the ICARUS detector capabilities in
electromagnetic shower reconstruction through the analysis of a
sample of events with hadronic interactions leading to the
production of $\pi^{0}$ mesons. The decay of a particle with known
mass gives a good test for the energy calibration. Another
motivation for such studies is the following: in order to
distinguish $\nu_{e}$ charge current (CC) interaction

\begin{equation}
\label{I1} \nu_{e} + n \rightarrow e^{-} + p,
\end{equation}

\noindent from the $\nu_{\mu}$ neutral current interaction

\begin{equation}
\label{I2} \nu_{\mu} + n \rightarrow \nu_{\mu} + \pi^{0} +
hadrons,
\end{equation}

\noindent a very good detector capability to separate electrons
and pions is required. The $\pi^{0}$ decays in about 98.8\% into
two photons. The Dalitz decays, leading to the production of one
photon and electron-positron pair, compose about 1.2\%. Both
photons and electrons may create electromagnetic showers and
therefore lead to the similar event topology for both NC and CC interactions.\\
The paper is organized as follows. The ICARUS T600 detector set-up
is briefly described in Section \ref{setup}. Section
\ref{selection} concerns the search and the selection of cosmic
ray hadronic interactions in T600 detector with the production of
the $\pi^{0}$ meson. In Section \ref{reconstruction} the methods
used for the measurement of the energy released in LAr by the two
electromagnetic showers coming from $\pi^{0}$ $\rightarrow$
$\gamma$ $\gamma$ decay and for the reconstruction of $\pi^{0}$
mass are illustrated. The analysis of the selected event sample
from real data is presented in Section \ref{realdata}. In Section
\ref{showres} the precision in the measurement of the
electromagnetic shower energy deposition in LAr is evaluated
in the condition of the test experiment through a simulation of
electrons and photons showering inside T600 detector. In Section
\ref{validation} the Monte Carlo simulations of $\pi^{0}$ in
ICARUS T600, used to validate the reconstruction routines for
electromagnetic showers, are shown. In Section \ref{nusim} the
same method of event reconstruction used for the real data is
applied to Monte Carlo simulated muon neutrino interactions with
$\pi^0$ production. The possible mixing of such events with the
electron neutrino CC interaction is discussed. The work is
summarized and concluded in Section \ref{conclu}.

\section{Experimental set-up}
\label{setup}

The T600 ICARUS detector is composed of two identical adjacent
"T300" half-modules filled with Liquid Argon. Two Time Projection
Chambers (TPCs) with common cathode placed in the middle, the
electric field shaping system, LAr purity monitors and
photo-multiplier tubes (PMTs) are enclosed in each half-module. The external
thermal insulation, together with a Liquid Nitrogen circuit,
allows to stabilize the LAr temperature. The required LAr purity
is achieved by a purification system. Each T300 half-module has
the following internal dimensions: 3.6 m (width), 3.9 m (height)
and 19.9 m (length). Along the longest side walls (left and right)
of each half-module, three parallel planes (two {\it Induction}
planes and one {\it Collection} plane) of anode wires are placed.
They are parallel to the cathode and form two TPCs. They are
oriented at 60$\deg$ with respect to each other, 3 mm apart, with
wire pitch of 3 mm. The total number of wires in the T600 detector
is 53248, whereas the volume of LAr in the instrumented part of
the detector is 340.3 m$^3$ (476.5 tons). The ionization
electrons, forced by a uniform electric field of 500 V/cm
perpendicular to the cathode, drift toward the anode wires inducing a
signal on those which they pass by. The appropriate voltage
biasing causes non-destructive signal to be induced first on two
Induction wire planes, whereas the ionization electrons charge is
finally collected in the Collection wire plane. By combining the
information from two-dimensional projections, provided by each
pair of wire planes, with the position along the drift time, it is
possible to obtain a three-dimensional reconstruction of an event.
The zero time $t_{0}$ of an event is defined by the PMT detection of
LAr scintillation light. The position along the
drift time is given with respect to $t_{0}$.\\
Several triggers have been applied during the T600 ICARUS detector
test run. An external trigger system made of plastic scintillators
was used for detection of muon tracks. An independent trigger
system was based on photo-multipliers, which detect scintillation
light produced during interaction of charged particles with LAr.
Most of the $\pi^{0}$ candidates used in this analysis were found in the data
collected with this internal trigger.
The detailed description of the T600 ICARUS detector design, construction,
and performance is given in~\cite{t600paper}, where
the conditions of the technical run and the event off-line
reconstruction procedure are discussed as well.

\section{Data selection}
\label{selection}

\begin{figure}
\centering
\begin{turn}{90}
\mbox{\epsfxsize=17.0cm\epsffile{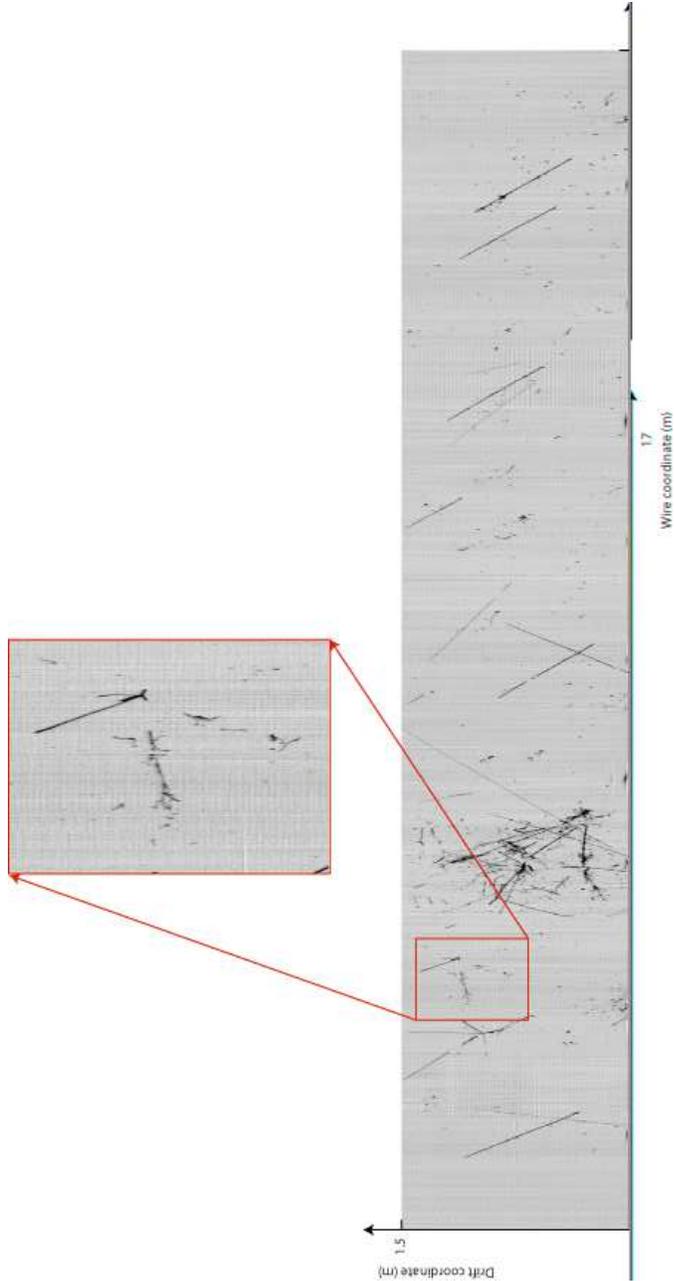}}
\end{turn}\caption {Example of an ICARUS T600 full view event
(Run 939 Event 130). On the left side, close to the center of the
image, a large hadronic interaction leading to the production of a
number of secondaries is visible. Along all the image m.i.p.
tracks can be seen. A zoomed view of an hadronic interaction
leading to $\pi^0$ production and subsequent decay in two photons
is shown. The scale of the image does not match the real one.}
\label{crowded}
\end{figure}

\begin{figure}
\centering \epsfig{file=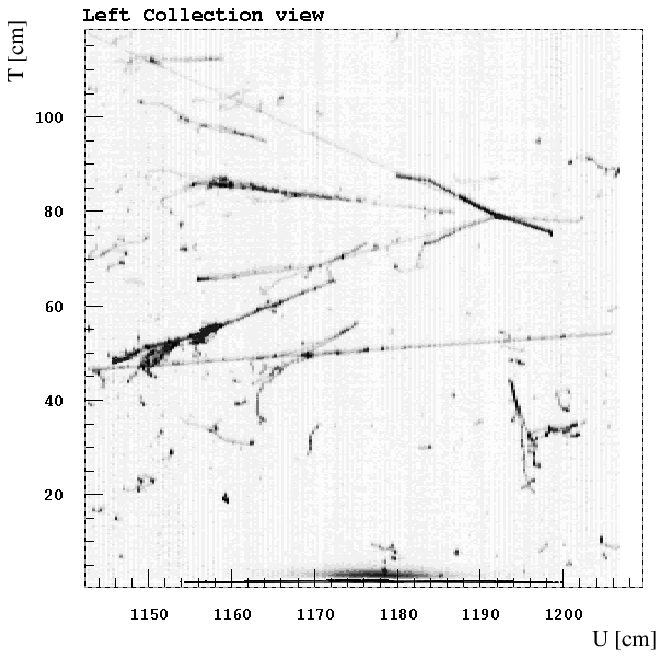,width=7cm,height=7cm} \caption {An
example of event in which the $t_0$ is uniquely determined from PMT
induced signal on Collection wires, visible at the drift coordinate
\emph{T}$\sim$ 0 cm as an ionization strip along the horizontal
\emph{U} wire coordinate (Run 700 Event 58).} \label{pmtinduced}
\end{figure}

Interactions in LAr producing $\pi^0$s decaying into photons have
been selected through a visual scanning for events with at least
two showers pointing to the interaction vertex: Fig. \ref{crowded}
shows an example of an ICARUS T600 full view event, with a $\pi^0$
candidate that was selected during the scanning phase. Most of the
selected interactions are initiated by hadrons, hence we also use
the term "hadronic interactions". Altogether, about 7500 images
have been scanned. The majority of the events were collected with
triggers based on signals from photo-multipliers, rich in hadronic
interactions.\\
It has to be stressed that the T600 ICARUS test run internal trigger based on
photomultipliers was aimed at air showers, so that many events were not
measurable because of their complexity due to a crowded environment.
The surface exposure has in fact as a result that a number of candidate $\pi^0$ events can lie in zones of
the detector of strong pile up of cosmic ray events, which can partially
overlap the electromagnetic showers of the photons of the $\pi^0$ decay. This affects either
the capability of the visual scanning in disentangling the good events from an environment populated
by many through-going particles or the precision in the shower energy
reconstruction for selected events with partial overlap with cosmic ray events.\\
The following scanning rules were adopted for the pre-selection of
hadronic interactions with $\pi^0$ production:

\begin{itemize}

\item At least two, well separated electromagnetic showers
originated by photons pointing to the interaction vertex are
demanded. Shower verteces and directions are defined by the
initial electron-positron pairs, so that it is often easy to
decide if they are pointing to the interaction vertex. The
showers, as well as the interaction vertex, must be clearly
visible in at least two views.\\

\item Knowledge of the event zero time ($t_0$) is mandatory in order
to determine the interaction vertex and the shower spatial
positions in the chamber, as well as to apply proper corrections
to the collected signals for the charge attenuation due to the finite
electron lifetime in Argon, associated to the electron attachment
process to electro-negative impurities.\\
A unique $t_0$ determination is possible in the following cases:
(1) at least one track (but not showers) from the interaction is
crossing the anode wires ($t_0$ = 0 $\mu$s for crossing points);
(2) at least one track (but not showers) is crossing the cathode
plane ($t_0$ = 962 $\mu$s for crossing points); (3) an unique
association between the interaction and
the PMT induces signals on Collection wires (Fig. \ref{pmtinduced}).\\

\item Interactions with neutral meson candidates are rejected
when ionization due to other tracks and interactions in their
neighborhood is so high that it seriously influences the shower
energy determination.

\end{itemize}

\begin{figure}
\centering \epsfig{file=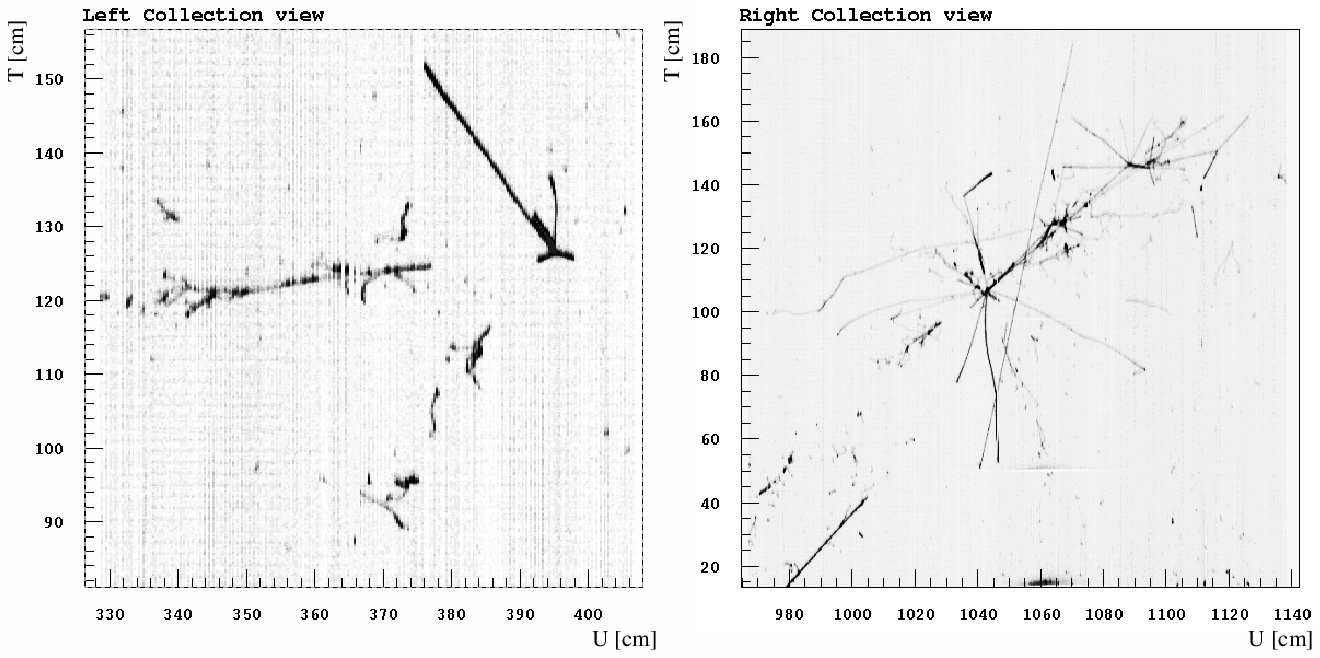,width=14cm,height=7cm}
 \caption {Examples of events from the final data sample: a) $\pi^0$
candidate (Run 939 Event 130); b) a multi-shower event (Run 939
Event 190).} \label{ev-accepted}
\end{figure}

\begin{figure}
\centering \epsfig{file=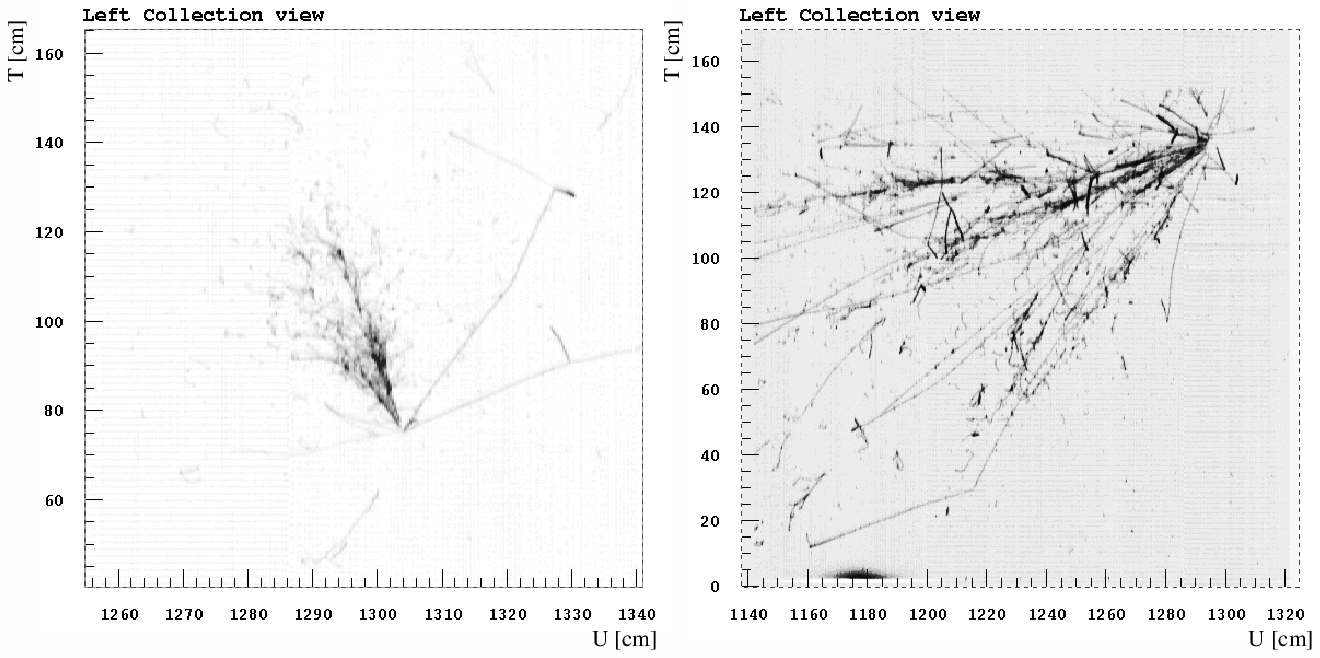,width=14cm,height=7cm}
\caption{Examples of candidates not included in the final data
sample: a) neutral meson candidate with overlapping photon
showers(Run 975 Event 65); b) candidate event in a crowded environment
(Run 710 Event 29).} \label{ev-rejected}
\end{figure}

\noindent Every candidate event was analyzed independently by (at least)
two groups from different laboratories in the Collaboration.
After all these requirements altogether 212 hadronic
interactions with at least one candidate neutral meson were
collected for further analysis. As said before, the crowded
environment of ICARUS T600 test run data reduced the usable data
sample substantially: examples of accepted and rejected events
with neutral meson candidates are shown in Fig. \ref{ev-accepted}
and in Fig. \ref{ev-rejected}, respectively.

\section{Data reconstruction}
\label{reconstruction}

\begin{figure}
\centering \epsfig{file=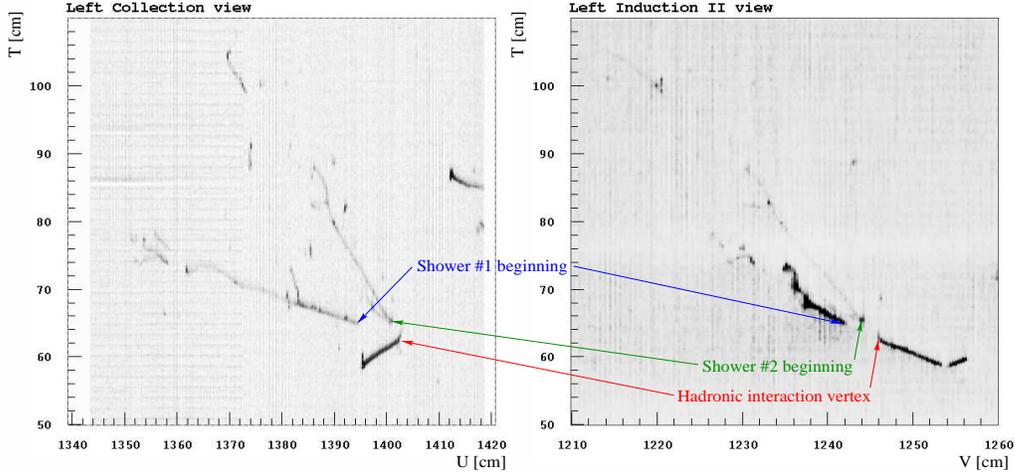,width=14cm,height=7cm}
\caption{Example of selected candidate $\pi^0$ $\rightarrow$
$\gamma$ $\gamma$ event (Run 712 event 7). Hadronic interaction
vertex and e.m. shower starting points are indicated in both
Collection and Induction views.} \label{selarea1}
\end{figure}

\begin{figure}
\centering \epsfig{file=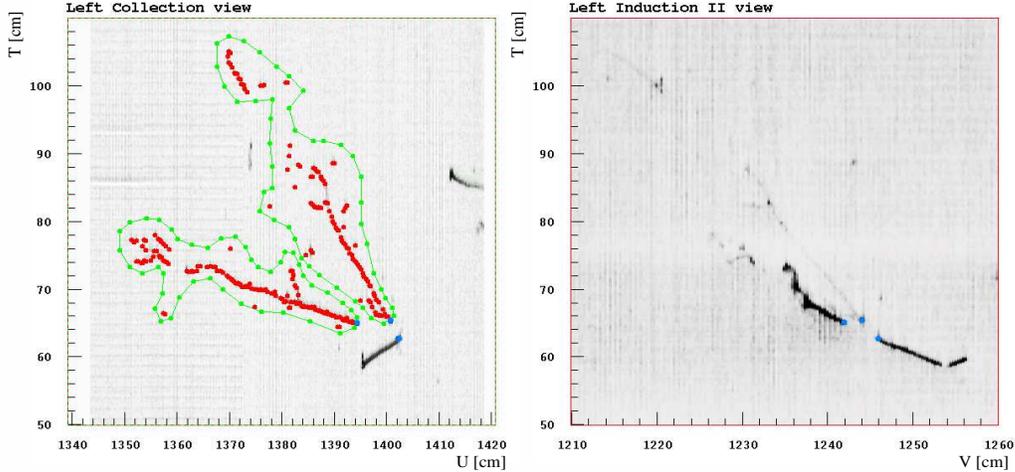,width=14cm,height=7cm}
\caption{Example of the application of the routines that allow to
select the hits belonging to an e.m. shower: the polygonal lines
are built by the user. Reconstructed hits inside the areas defined
by the polygonal lines are marked with red dots (Run 712 event
7).} \label{selarea2}
\end{figure}

The invariant mass $M$ of a particle decaying into two photons is
given by

\begin{equation}
M  = \sqrt{2~E_{1}E_{2}(1-cos\theta_{12})} \,, \label{pimass}
\end{equation}

\noindent where $E_{1}$ and $E_{2}$ are the energies of the two
photons and $\theta_{12}$ is the angle in between. In our case,
the goal was to measure the energies of two electromagnetic showers
coming from the photons and to reconstruct the photon directions.\\
In Fig. \ref{selarea1}, one of the 212 selected events is shown: in
both Collection and Induction views an hadronic interaction vertex
and the two shower starting points are visible. The drift position
coordinate \emph{T}, where $T$(cm) = \emph{drift time} (s) $\times$
\emph{electron drift velocity} (cm/s), is common for the two 2D
projections of the events, while the \emph{U}(cm) and \emph{V}(cm)
wire coordinates indicate the wire position in Collection and
Induction views, respectively. The \emph{U}, \emph{V} and \emph{T}
coordinates of the interaction vertex and those of the e.m. shower
starting points are used for the three dimensional reconstruction of
the photon direction, hence of the angle $\theta_{12}$ between
the two photons of the $\pi^0$ decay.\\
The basic step for the calorimetric reconstruction of
electromagnetic showers is the hit finding and reconstruction
procedure. We define $hit$ as the portion of particle track which
energy is read by a given wire. The hit search and reconstruction is
performed on the raw signal of the Collection wire plane following
the standard procedure for ICARUS data analysis described
in~\cite{t600paper}. After hit reconstruction, the following
parameters are available for each hit: its wire-drift position
\emph{U} and \emph{T} and
its area (ADC $\times$ $\mu$s).\\
A dedicated HIGZ-based Graphical User Interface visualization tool
allows to select directly the region with the hits belonging to a
shower, by building a polygonal line around the shower directly on
the Collection view by means of interactive calls of HIGZ ambient.
Then hit finding routine is applied within the bounds defined by
the polygonal line (see Fig. \ref{selarea2}). It has to be noticed
that the definition of the area surrounding the shower is left to
the user choice, so that small variations of the reconstructed
energy can be found between measurements done by different users.
This happens especially in the case the two photons of the $\pi^0$
decay are emitted at small angle, so it can be difficult to decide
whether an energy deposition belong to the first shower or to the
second one.\\ Once all hits belonging to a shower are found, the
overall shower energy $E$ is computed by summing all energies
$E_i$, where $i$ is the index of the hits inside the shower area:

\begin{equation}
E_i(MeV) = \frac{1}{\varepsilon} \frac{CW}{R}\cdot A \cdot Q_i \,, \label{energy}
\end{equation}

\noindent where: $C = (152 \pm 2) \times 10^{-4}$ fC/(ADC $\times$
$\mu$s) is the calibration factor \cite{purity};
$W = 23.6^{+0.5}_{-0.3}$ eV is the average energy needed for the
creation of an electron-ion pair \cite{ioni}; $R = 0.640 \pm
0.013$ is the electron-ion recombination factor \cite{muon_decay};
$A$ = $e^{\frac{t_i - t_{0}}{\tau_{e}}}$ is the drift electron lifetime
factor: $t_i$ is the hit drift time coordinate and $t_{0}$ is the
event zero-time. $\tau_{e}$ is the actual electron lifetime in
LAr, that varies run by run~\cite{purity}. Its value as a function of the
acquisition date was monitored during the T600 data
taking~\cite{recon}, with relative errors at the level of 5\%;
finally, $Q_i$ is the hit area. The shower
energy $E$ is then corrected by multiplying it by a factor
$\frac{1}{\varepsilon}$, coming from Monte Carlo simulations of
e.m. showers from photons and electrons in the range 50 $\div$
5000 MeV in ICARUS T600, taking into account the shower energy
reconstruction efficiency (see Sec. \ref{showres}). It has to be
noticed that the correction for the reconstruction efficiency in
eq. (\ref{energy}) depends on the attenuation factor $A$ =
$e^{-\frac{t_i - t_{0}}{\tau_{e}}}$ coming from the drift electron
finite lifetime of the Monte Carlo generated events, so that the
$\varepsilon$ value changes shower by shower, ranging from
$\varepsilon$ = 0.69 $\pm$ 0.03 ($A$ = 0.5) to $\varepsilon$
= 0.83 $\pm$ 0.04 ($A$ = 1) (see Fig. \ref{epsatt}). The error on the shower energy
is then evaluated from the propagation of all contributions
in eq. (\ref{energy}).\\
The angle $\theta_{12}$ between photons is then evaluated from the
spatial coordinates of hadronic interaction vertex and of starting
points of the showers as explained above.

\section{Results}
\label{realdata}

\subsection{($\gamma$, $\gamma$) invariant mass}

\begin{figure}[htbp]
\centering \epsfig{file=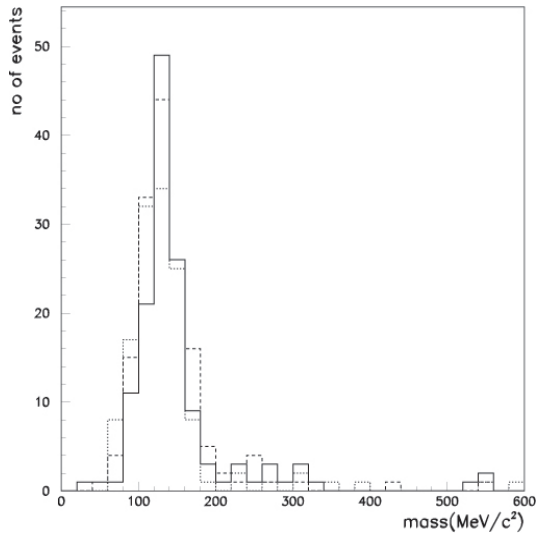,width=7cm,height=7cm}
\caption{($\gamma$,$\gamma$) invariant mass distributions (solid,
dotted and dashed lines) of three different laboratories involved
in the data analysis.} \label{threelab}
\end{figure}

\begin{figure}[htbp]
\centering \epsfig{file=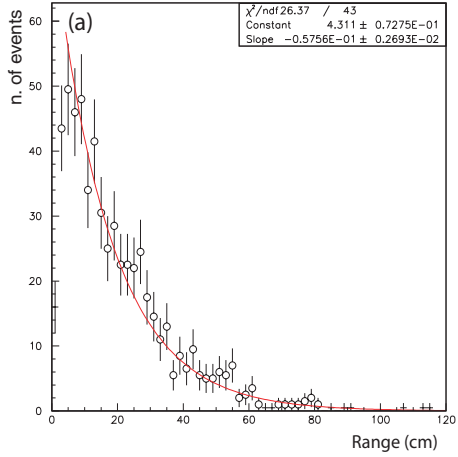,width=6cm,height=6cm}
\caption{Distribution of the distances between shower starting
point and interaction vertex for LAr radiation length
evaluation.} \label{radlen}
\end{figure}

\begin{figure}[htbp]
\centering \epsfig{file=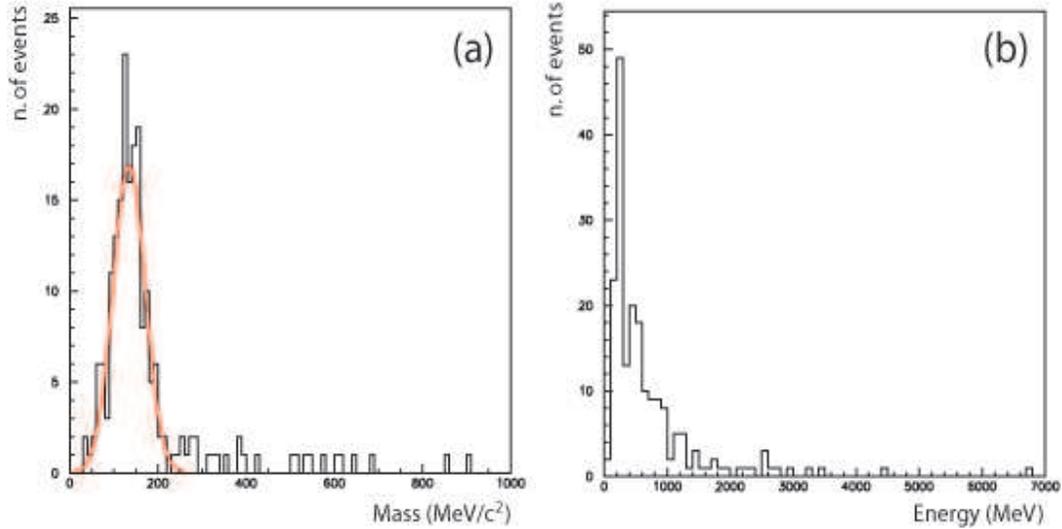,width=14cm,height=7cm}
\caption{($\gamma$,$\gamma$) invariant mass (a) and energy (b)
distributions for the sample of 196 $\pi^0$ candidates. The average
mass is $m_{\gamma\gamma}$ = 134.4 $\pm$ 3.0 MeV/c$^2$ ($\sigma$ = 36.7 $\pm$ 3.4 MeV/c$^2$).
The mean value of
the $\pi^0$ energy is \emph{E} = 707.7 MeV.} \label{allmass}
\end{figure}

Most of the 212 events forming the sample were measured in several
laboratories. The difficult condition of data analysis due to the
presence of a crowded environment sometimes has as a result that
the same event measured in different laboratories produces
different values in vertex coordinate and shower energy
determination. As explained above, the somehow arbitrary choice
of the user to build the polygonal line surrounding the e.m.
photon shower of the $\pi^{0}$ decay could in principle
introduce some systematics in the ($\gamma$,$\gamma$) invariant
mass distributions coming from different laboratories. Fig.
\ref{threelab} shows the ($\gamma$,$\gamma$) invariant mass
distribution of three sub-samples of events, each provided by a
different laboratory involved in the data analysis. It appears
that on average no systematic differences between measurements
carried out in different laboratories are present. At this stage
of the analysis, the $\varepsilon$ parameter taking into account the shower energy
reconstruction efficiency, see eq. (\ref{energy}), has not been considered yet, hence the
mass values are underestimated with respect to the nominal ones.
As a remark, it has to be noticed that in LNGS underground condition the events
will be very clean, so that each $\pi^0$ decay event will be
measured just once unambiguously. The event sample for the
following analysis was then built by taking just one
measurement per event chosen randomly from the available ones in
the case of events measured by more than one laboratory, then
adding to the sample all the other events measured just once.\\
A preliminary consistency check was performed on this event
sample: the distribution of distances of the shower starting
points from the corresponding interaction vertex is related to the
radiation length $X_0$ in LAr, $X_0$ = 0.14 m \cite{pdg}. This is
shown in Fig. \ref{radlen}. The value of the mean distance
obtained from a fit with an exponential function is:

\begin{equation}\label{eqradlen}
    X_\gamma = (0.174 \pm 0.008) \, m \,,
\end{equation}

\noindent that is in agreement with the expected length for
photons $X_\gamma$ = $\frac{9}{7}$ $\cdot$ $X_0$ = 0.18 m for
LAr.\\ A fiducial volume cut was then applied by requiring that at
least the first three radiation lengths of both showers were fully
contained in the chamber. This reduced the event number to 164, of
which 126 were two-shower and 38 were multi-shower events; the
final number of $\pi^{0}$ candidates in the selected event sample
was then found to be equal to
196. The invariant ($\gamma$,$\gamma$) mass distribution of this
196 $\pi^{0}$ candidate sample is shown in Fig.~\ref{allmass}, as
well as the $\pi^{0}$ energy distribution. The correct mass
combinations of multi-shower events were disentangled after the
minimization of the parameter:

\begin{equation}
     \chi^2 = \sum_{i=1}^{n} (\frac{m_{\pi^{0}} -
     m_{i}}{\sigma_{i}})^2
     \label{chi2}
\end{equation}

\noindent evaluated for all the possible combinations of showers.
($m_{\pi^{0}}$ = 135 MeV/c$^2$~\cite{pdg}).
Here, \emph{n} is the maximum number of $\pi^{0}$s allowed (1 for
a three shower event, 2 for a four and five shower event, 3 for a
6 shower event). The average mass given by a gaussian fit is:

\begin{equation}
    m_{\gamma\gamma} = 134.4 \pm 3.0 MeV/c^2 \,.
    \label{rawmass}
\end{equation}

\noindent The evaluation of all systematic contributions to the error on the $\po$ mass
(see Sec. \ref{reconstruction}) introduces an additional contribution to the mass
error of 7.1\% (9.5 MeV/c$^2$). The RMS uncertainty of the reconstructed mass was
found to be equal to $\sigma$ = 36.7 $\pm$ 3.4 MeV/c$^2$ (27.3\%). Thus the result,
in agreement with the $\po$ mass, demonstrates that our mass reconstruction
method is reliable.

\subsection{Crowded environment suppression}

\begin{figure}[htbp]
\centering \epsfig{file=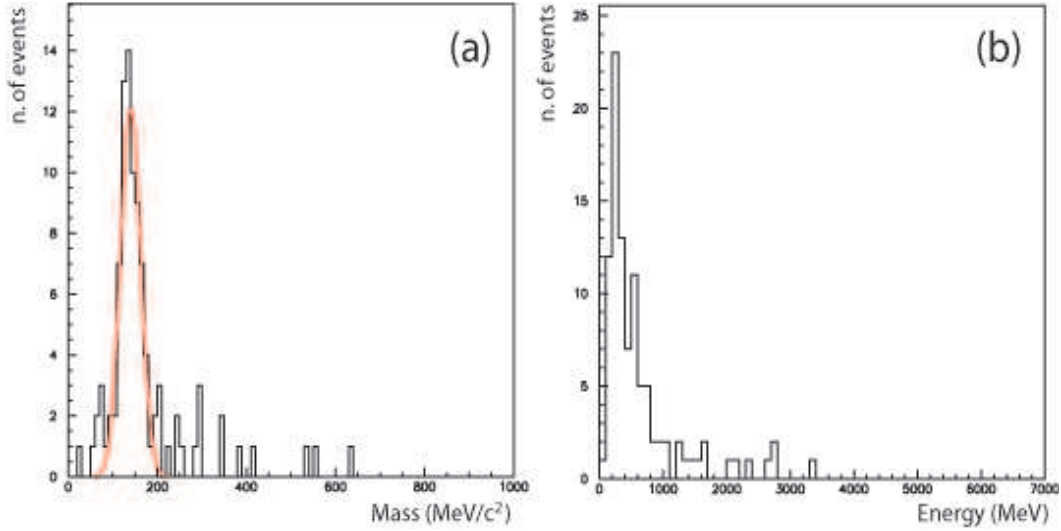,width=14cm,height=7cm}
\caption{($\gamma$,$\gamma$) invariant mass (a) and energy (b)
distributions for the final sample of 97 $\pi^0$ candidates
obtained after crowded environment suppression. The average mass
is $m_{\gamma\gamma}$ = 139.9 $\pm$ 2.8 MeV/c$^2$ ($\sigma$ = 22.6 $\pm$ 2.9 MeV/c$^2$).
The mean value of
the $\pi^0$ energy is \emph{E} = 641.8 MeV.} \label{twomass}
\end{figure}

Although on average the measurements done in different
laboratories are in agreement, see Fig. \ref{threelab},
nonetheless for this particular analysis of events collected
on surface by exposure by cosmic rays it has
been chosen to improve the results by cleaning the 212 event
sample from the events measured in a crowded environment. To do
so, further rejection cuts were related to the consistency of the
two independent measurements performed in different laboratories.
The purpose of this refined analysis is to show that in an environment
more similar to the one expected in the underground laboratory
the $\pi^0$ mass distribution has to be narrower than the one found for the
full sample.\\
The interaction vertex and shower starting point coordinates are
then considered to be in agreement if the absolute value of the
differences between two independent measurements of all three
coordinates is smaller than five times the width of the
distribution of such differences. Then, if the relative difference
$\frac{(E_1 - E_2)}{\frac{(E_1 + E_2)}{2}}$ between measurements
of the shower energies is smaller than 50\%, then the measurements
are considered to be in agreement, so that only these are used to
built the final $\pi^0$ mass distribution. The values used for
these cuts largely suppress the $\pi^0$ decay events belonging to
a crowded environment while keeping a relatively high statistics
for the following analysis. For events which passed these tests
the average values of the measured positions and energies were
calculated and used for $\pi^{0}$ mass determination.\\
The sample which passed these selection criteria was then
composed of 89 events with at least one $\po$
candidate; 72 were two-shower and 17 were multi-shower events. As
done for the global data sample, the correct mass combinations of
multi-shower events were disentangled after the minimization of
the $\chi^2$ parameter, see eq. (\ref{chi2}), evaluated for all
the possible combinations of showers. After this procedure, a
final sample of 97 $\pi^{0}$ candidates was available for the
analysis.\\ The ($\gamma$,$\gamma$) invariant mass distribution
for the final sample of 97 $\pi^{0}$ candidates is shown in
Fig.~\ref{twomass}a together with the total $\po$ energy in
Fig.~\ref{twomass}b. The average mass given by a gaussian fit was:

\begin{equation}\label{eqtwomass}
    m_{\gamma\gamma} = 139.9 \pm 2.8 MeV/c^2 \,,
\end{equation}

with a systematic contribution of 9.9 MeV/c$^2$.
The $\po$ mass resolution is now equal to 16.1\% ($\sigma$ =
22.6 $\pm$ 2.9 MeV/c$^2$), that is 8.9\% less than the one
evaluated for the sample without consistency cuts. This demonstrates
that in the particular case of the ICARUS T600 Earth surface data,
a refined analysis was necessary in order to take into account the
strong contribution of the crowded environment surrounding some of
the $\pi^{0}$ production and decay events. These crowded events are
badly measured and bring to major differences between measurements
carried out in different laboratories.

\section{Energy resolution for electromagnetic showers}
\label{showres}

\begin{figure}[htbp]
\centering \mbox{\epsfxsize=7.0cm\epsffile{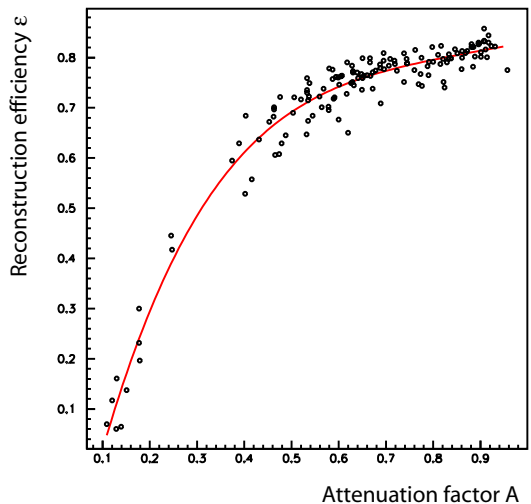}}
\caption{Reconstruction efficiency $\varepsilon$ as a function of
drift electron finite lifetime factor $A$: data have been fitted
with a fourth degree polynomial.} \label{epsatt}
\end{figure}

\begin{figure}[htbp]
\centering \mbox{\epsfxsize=7.0cm\epsffile{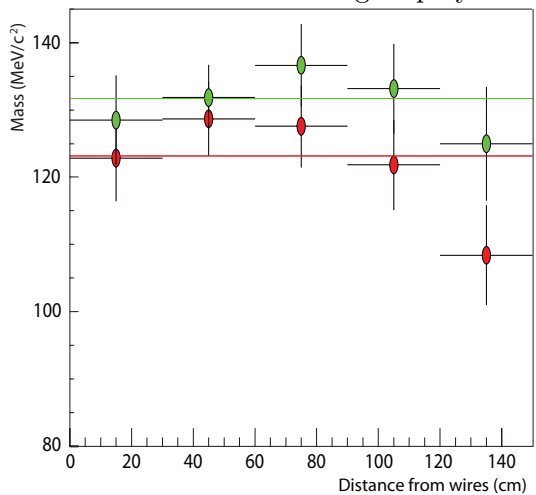}}
\caption{Dependence of the reconstructed $\pi^0$ mass from the
distance between the main hadronic interaction vertex and the wires
with (green dots) and without (red dots) the correction for
the $\varepsilon$ reconstruction efficiency parameter. The red (green) straight
represents the result of a fit of the red (green) dots with a flat line.} \label{my}
\end{figure}

In this section an analysis of the energy resolution for
electromagnetic showers in the energy range 50 $\div$ 5000 MeV is
presented. The energy of Monte Carlo simulated electrons and photons
inside the ICARUS T600 detector was reconstructed with the same
tools used for the $\pi ^{0}$ events. The
simulation was performed using the FLUKA~\cite{fluka} program for the
samples of electrons and photons generated at six different energies
with different orientation in the detector. The measured average
energies appeared to be shifted with respect to the true energies
defined at the generation. To obtain an agreement between
generated and reconstructed energies, the shower energy
reconstruction efficiency (i.e. the ratio between measured and
generated shower energy) $\varepsilon$ has to be taken into account.
The inefficiency can be due to misidentification of the hits by the
hit finding algorithms, as well as to the loss of the tail of the
reconstructed tracks, whose energy deposition can become lower than the
intrinsic wire background noise, especially in the case of low
values of the electron lifetime $\tau_{e}$: in this case the drift
electron finite lifetime factor $A$ = $e^{-\frac{t_i -
t_{0}}{\tau_{e}}}$ is crucial. In Fig. \ref{epsatt} the
reconstruction efficiency $\varepsilon$ as a function of $A$ is
shown, measured for a sample of Monte Carlo events at several values
of $\tau_{e}$: it is found that $\varepsilon$ rises with $A$, and
that it is well fitted with a fourth degree polynomial. Despite the
arbitrariness of the choice of this particular function for the fit of $\varepsilon$ versus $A$,
it was found that choosing other types of curves for the fit did not introduce systematics in the
value of $\varepsilon$, the fluctuations of which remain always well below the 1\% in the region
$A$ = (0.5$\div$1.0) where the real data lie.\\
Depending on the values of this attenuation factor measured in the real data,
$\varepsilon$ was found to range from 0.69 $\pm$ 0.03 ($A$ = 0.5) to
0.83 $\pm$ 0.04 ($A$ = 1). Moreover, in the range of e.m. shower
energy analyzed, no dependence of $\varepsilon$ from the shower
energy, the attenuation factor being the same, was found. The
effect of the $\varepsilon$ parameter taking into account the shower
energy reconstruction efficiency is shown in Fig. \ref{my}, where
the dependence of the reconstructed $\pi^0$ mass from the distance
between the main hadronic interaction vertex and the wires is shown
(sample of the 196 $\pi^0$ candidates). When the $\varepsilon$
efficiency is not taken into account (red dots),
an effect of underestimation of the mass, especially near the cathode ($y$ = 150
cm), is evident. After the shower energy correction (green dots), this effect
is much smaller and, as expected, the correction is larger for events
far from wires ($y$ $\gtrsim$ 60 cm). This effect can be seen in a more quantitative
way by giving the values of the $\chi^2$s from the fit of the two distributions with a flat line.
The value of $\chi^2/ndf$ after the shower energy correction ($\chi^2/ndf$ = 1.55/4)
is significantly smaller than the one without correction ($\chi^2/ndf$ = 5.52/4), demonstrating
that after the correction the mass distribution is almost flat.

\begin{figure}[htbp]
\centering \epsfig{file=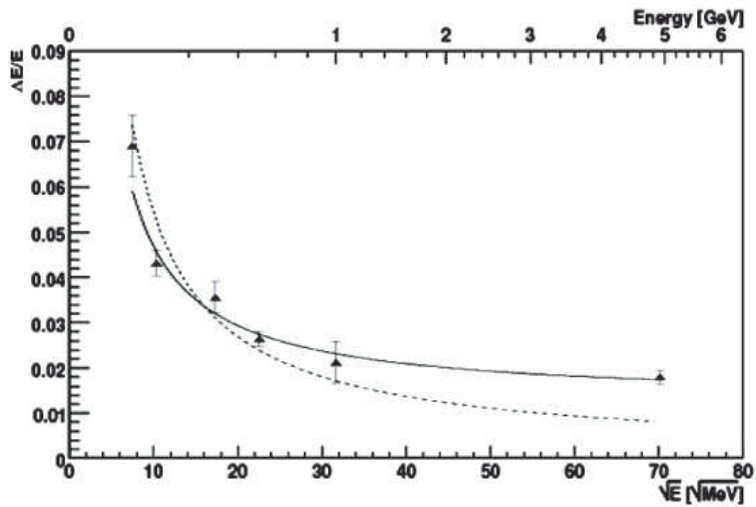,width=10cm,height=8cm}
\caption{\label{enerres} $\frac{\Delta E}{E}$ as a function of
$\sqrt{E}$. Dashed line - fit to eq. (\ref{eres2}) (P2 = 0
assumed), solid line - fit to eq. (\ref{eres2}).}
\end{figure}

\noindent The energy resolution at fixed energies is given by the
RMS of the measurements. The following formula was fitted to the
experimental points:

\begin{equation}
\label{eres2}
 \frac{\Delta E}{E} = \frac{P1}{\sqrt{E}} + P2
\end{equation}

\noindent where $P1$ and $P2$ are the free parameters of the fit.
The first part of the formula represents the typical
electromagnetic calorimeter behavior. The  introduction of $P2$
parameter improves the fit quality. Its presence can be due to the
fact that the amount of low energy signals at the end of the
showering process grows with the initial energy. These signals can
be mixed with the overall background when the hits belonging to
the shower are selected. Fig. \ref{enerres} illustrates the
resulting energy resolution of the T600 ICARUS detector in the
reconstruction of the electromagnetic showers.

\noindent The best fit with the formula of eq. (\ref{eres2}) was
obtained for the following parameter values:  $P1$ = (0.33 $\pm$
0.03) $\sqrt{MeV}$ and $P2$ = 0.012 $\pm$ 0.002. The energy resolution
in LAr detector was previously studied on a base
of a sample of MC generated low energy electrons (E $<$ 50MeV)~\cite{muon_decay}:
the best fit was obtained for $P1$ = (0.11 $\pm$ 0.01) $\sqrt{MeV}$ and $P2$ = 0.025 $\pm$ 0.03.
However in that case $P1$ and $P2$ were correlated because of the small energy
interval involved.
The worse resolution for the high energy electromagnetic showers can be explained by
larger fluctuations of integrated background because of the larger
area covered by the shower as well as to the hits overlapping in
the regions of large density of electron tracks. Moreover, in the
case of a high energy shower it is not obvious which surrounding
signals still belong to the shower. The energy resolution has been
obtained for the specific method of energy reconstruction based on
calorimetric information from Collection view only.\\
As a check of the reliability of the energy reconstruction method, the
distribution of the so called $pulls$ :
$pull(m_{\pi^0})=\frac{(m_{\gamma\gamma}-m_{\pi^0})}{\sigma(m_{\gamma\gamma})}$
was calculated for the real data, using eq. (\ref{eres2}) to
calculate the error $\sigma(m_{\gamma\gamma})$ as a function of
$\sigma(E_{\gamma})$. The RMS of the distribution was found to be
equal to one within errors, confirming the validity of the formula
obtained on the base of the measurements of the photon showers.

\section{Monte Carlo simulation of $\pi^0$ production from hadronic interactions}
\label{validation}

Monte Carlo simulations of three different samples of $\pi^0$s
with different kinematic and event environment conditions were carried out.
The events have been generated using the FLUKA
full simulation of T600 detector. They allowed to confirm the
validity of the $\pi^0$ mass reconstruction tools in environments
similar to the one coming from real data. The three generated
$\pi^0$ samples were:

\begin{enumerate}

\item $\pi^0$ decays at rest; the invariant mass is reconstructed
as a sum of the photon energies
 since the relative angle is exactly equal to 180$\deg$. Only
 errors on the photon energies contribute to the mass
 resolution.\\

\item $\Sigma^+ \longrightarrow \pi^0 + p$ with subsequent $\pi^0$
decay; the pion from the two body decay at rest has fixed momentum
of 189 MeV/c and the vertex of $\pi^0$ decay is precisely
determined by the position of the low energy proton from
$\Sigma^+$ decay. Shower verteces are also well determined because
no tracks overlap and affect their recognition.\\

\item $\pi^0$ decays observed in interactions of $\pi^-$ mesons
with Argon nuclei at 2 GeV incident energy. The events resemble
most our real events, corresponding to hadronic interaction.

\end{enumerate}

\begin{figure}[htbp]
\hspace*{-1.5cm} \epsfig{file=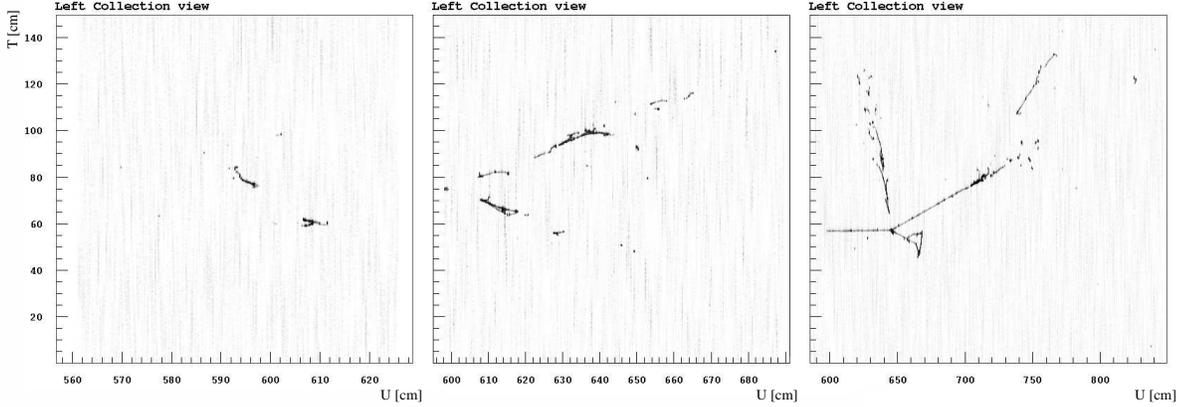,width=16cm,height=6cm}
\caption{Examples of neutral pion decays as seen by FLUKA-ICARUS
detector: decay at rest (a), $\pi^0$ from $\Sigma^+$ decay (b) and
from $\pi^-$ interaction in LAr (c).} \label{mcevents}
\end{figure}

\begin{figure}[htbp]
\hspace*{-2cm}
\epsfig{file=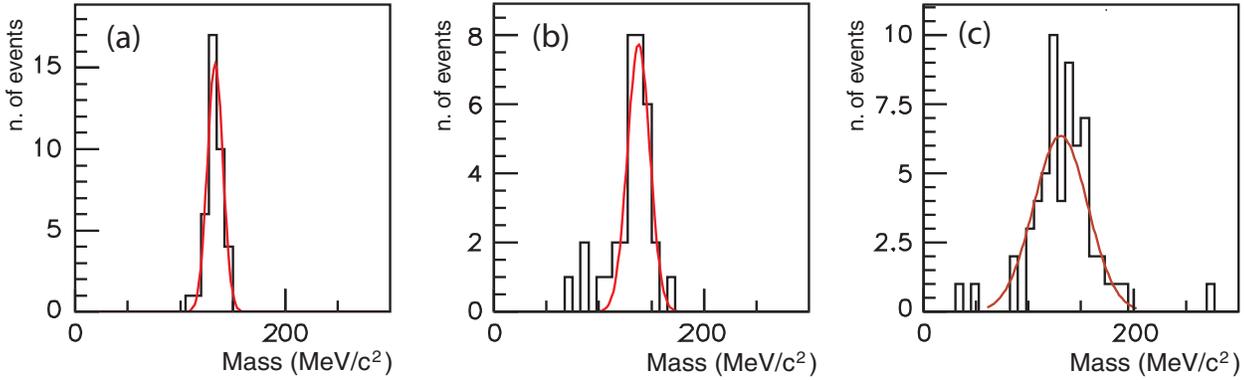,width=16.5cm,height=5cm}
\caption{(a) $m_{\gamma\gamma}$ for pion decays at rest:
$m_{\gamma\gamma}$ = 133.3 MeV/c$^2$ (RMS = 7.4 MeV/c$^2$); (b)
$m_{\gamma\gamma}$ from $\Sigma^+$ $\longrightarrow$ $\pi^0 + p$
decay: $m_{\gamma\gamma}$ = 137.8 MeV/c$^2$ (RMS = 10.7
MeV/c$^2$); (c) $m_{\gamma\gamma}$ from 2 GeV $\pi^-$ LAr
interactions: $m_{\gamma\gamma}$ = 130.6 MeV/c$^2$ (RMS = 25.8
MeV/c$^2$).} \label{mcmass}
\end{figure}

\noindent Examples of each class of these events are shown in Fig.
\ref{mcevents}. The ($\gamma$,$\gamma$) invariant mass
distribution for the three measured Monte Carlo samples is shown
in Fig.~\ref{mcmass}a, \ref{mcmass}b and \ref{mcmass}c. It can be
seen that the position of the $m_{\gamma\gamma}$ peak is
consistent with the pion mass for all three generated $\pi^{0}$ samples.\\
The RMS of the invariant mass distributions presented in Fig.
\ref{mcmass} are equal to (7.4 $\pm$ 1.3) MeV, (10.7 $\pm$ 4.8)
MeV and (25.8 $\pm$ 4.7) MeV. The evident increase in RMS is
mainly due to the event crowded environment, which augments the
difficulty of the event interpretation because of the higher
energy and the consequent presence of other reaction products.
Moreover, in the decay at rest the relative angle between photons
was not measured, but assumed to be equal to 180$\deg$. It has to
be noticed that Monte Carlo events were generated in the center of
the detector. For real events the effect of some energy leakage
through the walls of the detector can appear despite of the
fiducial volume cut.

\section{Simulation of $\po$ production in $\nu$ interactions}
\label{nusim}

\begin{figure}[htbp]
\centering \mbox{\epsfxsize=14.0cm\epsffile{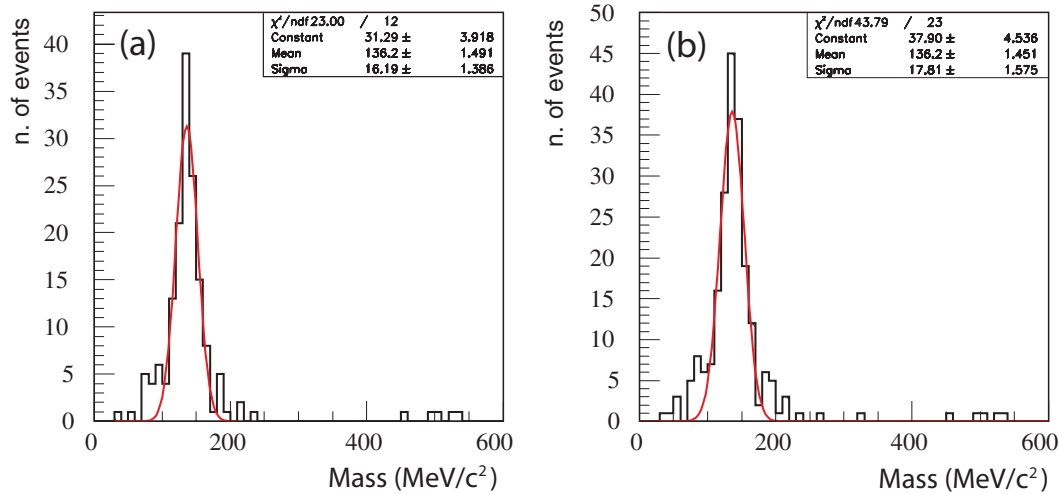}}
\caption{ \label{pi0_mc_all_mgg} Two photon invariant mass from
events initiated by muon neutrino NC interactions. Pions from both
primary and secondary verteces are included; one pion events (a),
all (one and multi-pions) events (b).}
\end{figure}

In the previous sections it has been shown that the results of the
analysis of real and simulated events with $\po$ production are
consistent. Therefore, the same method of generation, measurement
and reconstruction was applied to simulated neutrino-induced pion
production to check the potentiality of the LAr technique
in this respect. An important point in the future experiments
looking for $\theta_{13}$ is the efficiency of distinction between
photon-induced showers from $\pi^0$ decay and the electron showers
from electron neutrino interactions.\\
The neutral pion production from NC neutrino interactions in the
Liquid Argon ICARUS detector was simulated using NUX-FLUKA chain
of programs. The neutrino energies  were chosen in a region of
maximal oscillation probability between 1 and 5 GeV.\\
The measured distribution of the two photons invariant mass is presented in
Fig. \ref{pi0_mc_all_mgg}. To disentangle the proper photon pair
combination for events with four and six photons pointing to the
same vertex, the pairs with minimum $\chi^2$ of the difference
between measured values of ($\gamma$,$\gamma$) invariant mass and
the $\po$ mass were selected, see eq. (\ref{chi2}). The position
of the maximum was found to be consistent with the $\po$ mass
value. The emission of 5 $\eta(547)$ mesons was observed in the
generated sample and its mass was properly reproduced. The width
of the $\pi^0$ peak is $\sigma$ = 17.8 $\pm$ 1.6 MeV/c$^2$
(13.1\%). This value is similar to the one obtained for the real data
($\sigma$ = 22.6 $\pm$ 2.9 MeV/c$^2$, 16.1\%), the difference
being probably due to the larger background at the earth surface.
The shape of $m_{\gamma\gamma}$ distribution presents deviations
from the gaussian shape as it was found with the real data.\\ After
the inspection of Monte Carlo events coming from the tails of the
two photons invariant mass distribution it was found that some
$\po$s were emitted from a secondary neutral vertex close to the
primary one. Some of them were the products of $\Lambda$ or K
meson decays. Wrong assignment of the decay vertex obviously leads to a
shift in the reconstructed mass. In the LAr TPC the good spatial
resolution allows to distinguish majority of the secondary verteces if their distance
from the main hadronic interaction vertex is larger than about 1 cm. The problem is
much more severe in water or sandwich-like
detectors which are characterized by much worse spatial resolution.
It should be noted that the expected percentage of the all secondary
neutral pions was found to be as large as  about 25\% at 2 GeV and  43\% at 5 GeV in FLUKA based generator.

\subsection{Neutral current $\nu _\mu$ events as a background to $\nu_e$ CC interactions}

An event with $\po$ production in NC  $\nu _\mu$ interaction can
be misidentified with an electron from CC $\nu_e$ interaction if
one of the two following configurations occurs:

\begin{itemize}

\item[1)] one photon from the decay can be lost when its energy
it is too small to allow its distinction in the crowded environment surrounding it (E $<$ 20 MeV);
the calculation based on the decay kinematics shows that the percentage
of events with such decay photon, being a function of the parent pion energy,
never exceeds 2.5\%. The showers of energy larger than 20 MeV were always properly recognized
in the simulated events.\\

\item[2)] The two photon showers overlap:
the angle between the two photon showers from the decay is sufficiently large to allow to distinguish the two showers
only for a $\pi^0$ energy up to about 1 GeV. The angle decreases from 70.1$\deg$ at 100 MeV to 13.7$\deg$ at 1 GeV.
If one does not require a full $\pi^0$ mass reconstruction as a pion selection tool the shower overlap is less critical. The possible overlapping effect for more energetic pion decays was already studied: it was shown~\cite{dedx} that the
$\frac{dE}{dx}$ method can be efficiently applied, in LAr detector,
to distinguish the beginning of the electron track from the
beginning of the lepton pair that comes from the photon
conversion.
\end{itemize}

Moreover, internal conversion due to the Dalitz decay (1.2\% of cases) or conversion
close to the primary vertex (5.5\% of cases within the first 1 cm) gives raise to misidentification
of a photon shower as an electron induced cascade in the 6.7\% of cases. In the most frequently populated interval of $\po$ energy (200$\div$400 MeV), in about 2\% of events the low energy photon can be lost. By multiplying this probability by 0.067
it follows that, in our detector, the total probability of the $\pi^0$ background is expected to be
at the level of 10$^{-3}$.

\section{Conclusions}
\label{conclu}

A sample of 212 events with at least two photon showers pointing
at the apparent interaction vertex were selected from the data
collected during the surface test run  of the ICARUS detector. Methods of
measuring the shower energy and shower direction were developed and
the invariant mass of the photon pairs was reconstructed. The
position of the maximum in the invariant ($\gamma$,$\gamma$) mass
distribution for the 196 $\pi^0$ candidates, obtained after a fiducial volume
cut, was found to be in agreement with the value of the $\pi^0$ mass,
thus confirming the common assumption that the $\pi^0$s are the main
source of the photons. The RMS uncertainty of the
reconstructed $\pi^0$ invariant mass was found
to be equal to 36.7 $\pm$ 3.4 MeV/c$^2$ (27.3\%).\\
An improved analysis was carried out in order to clean the initial
event sample from the events measured in a crowded environment
caused by other tracks and interactions surrounding the
$\pi^0$ candidates. As a result, the position of the maximum in
the invariant ($\gamma$,$\gamma$) mass for the 97 surviving
$\pi^0$ candidates was found again to be in agreement with
the $\pi^0$ mass. Moreover, the RMS uncertainty of the reconstructed
$\pi^0$ invariant mass was found to be equal to (22.6 $\pm$ 2.9)
MeV/c$^2$ (16.1\%), which is significantly narrower than the
mass distribution for the initial event sample. The $\pi^0$
mass distribution was in good agreement with
the distributions obtained for the Monte Carlo generated events with
neutral pion production in hadronic and neutrino interactions.
The agreement between the data and the Monte Carlo confirms the validity
of the shower energy and shower direction reconstruction methods.
It also demonstrates the potential of the ICARUS LAr TPC, because
simple quality cuts were sufficient to obtain good results in very
difficult conditions of the surface test, detector commissioning
and tuning of the electronics.\\
It was found that in the photon energy range from 50 to 5000 MeV
the electromagnetic shower energy resolution $\sigma(E)/E$ can be
approximated by the formula $\sigma(E)/E$ = $P1/\sqrt{E}$ + $P2$ with
the parameters $P1$ = (0.33 $\pm$ 0.03) $\sqrt{MeV}$ and $P2$ = 0.012 $\pm$ 0.002.\\
To simulate the future measurements in the CNGS beam, neutrino
interactions at incident energies between 1 and 5 GeV were generated using the
FLUKA package. Scanning and measurements of the showers pointing
to the primary and secondary verteces were carried out, and the two
photon mass was reconstructed for events with different number of
showers. It was demonstrated that in most cases a proper
reconstruction of up to the three pions coming from the same
vertex was possible. A few events with $\eta(547)$ meson decay into two photons were
found in simulated events and  properly reconstructed.\\
The asymmetric decays of $\pi^0$s as a background to $\nu_e$ charge
current events were discussed. It was found that it is likely to
loose the photon of energy smaller than about 20 MeV but the percentage
of events with such a photon never exceeds 2.5\%. The frequency of events
that could be misidentified with events containing single
electron shower, because of the low energy photon lost for either
the Dalitz decay or the events with the second
photon converting close to the neutrino interaction vertex
was estimated to be at the level of one per mil. \\
In summary: our analysis implies very good prospects for the
recognition and energy measurement of electromagnetic showers
during the ICARUS running in the Gran Sasso laboratory,
where the background due to other signals in the
neighborhood of neutrino interactions will be negligible.

\section*{Acknowledgements}
We wish to dedicate this work to the memory of our friend and colleague
Fulvio Mauri, co-author of the paper, who prematurely passed away on March, 2008.\\
We would like to warmly thank the many technical collaborators
that contributed to the construction of the detector and to its
operation. We are glad of the financial and technical support of
our funding agencies and in particular of the Istituto Nazionale
di Fisica Nucleare (INFN). The Polish groups acknowledge the support
of the Ministry of Science and Higher Education in Poland,
105,160,620,621/E-344,E-340,E-77,E-78/SPB/ICARUS/P-03/DZ211-214/2003-2008,
1P03B04130 and N N202 0299 33.

\end{document}